\documentclass[11pt,twoside]{article}

\usepackage{asp2006}
\usepackage{graphicx}

\begin{document}
\title{Giant Planet Formation by Core Accretion}   
\author{Christoph Mordasini, Yann Alibert and Willy Benz}   
\affil{Physikalisches Institut, Sidlerstrasse 5, CH-3012 Bern, Switzerland}    
\author{Dominique Naef}   
\affil{ESO, Alonso de Cordova 3107, Casilla 19001, Santiago 19, Chile}    
\begin{abstract} 
We present a review of the standard paradigm for giant planet formation, the core accretion theory. After an overview of the basic concepts of this model, results of the original implementation are discussed. Then, recent improvements and extensions, like the inclusion of planetary migration and the resulting effects are discussed. It is shown that these improvement solve the ``timescale problem''. Finally, it is shown that by means of generating synthetic populations of (extrasolar) planets, core accretion models are able to reproduce in a statistically significant way the actually observed planetary population. 
\end{abstract}

\section{Introduction} 
Our current understanding of planet formation is based on several centuries of observations of the planets of our own Solar System, 12 years of extrasolar planets detection,  and several decades of observations of young stellar systems. These studies have let to the general concept that after the collapse of a dense gas cloud, a protostar surrounded by a protoplanetary disk was formed. In this disk, solids started to coagulate from fine dust and grew further by mutual collision to form planetesimals (provided the bottleneck by bodies roughly one meter in size can be overcome), then protoplanets, and ultimately the actual planets. Some of the protoplanets managed to accrete a massive gaseous envelope onto their core. This is the very rough outline of the core accretion model.

\section{Core Accretion Paradigm}
In the core accretion model, the formation of gas giant planets is therefore seen as a two step process: In the first one a solid core is formed which, provided this core reaches critical mass, gives rise to the second step in which runaway gas accretion occurs, leading to a quick build-up of a massive  envelope. This basic scenario was first studied more than 30 years ago \citep{CMperricameron1974, CMmizunoetal1978,  CMbodenheimerpollack1986}.

\subsection{Baseline Formation Models}
The growth of the core (at a fixed semimajor axis in baseline models) occurs through collisional accretion of background planetesimals, which themselves are formed by collisional coagulation of small dust grains \citep{CMwetherillstewart1989} or an instability in the dust layer \citep{CMjohansen2007}. The governing equation for the growth rate is given by the classical theory of \citet{CMsafronov1969}. However, in many early studies, $dM_{\mathrm{core}}/dt$ was treated as a free parameter. The formation of the core occurs through the same mechanism as the one generally accepted for the formation of terrestrial planets. In this sense, core accretion is an unified theory for the formation of both terrestrial and giant planets.  When the core has reached roughly the mass of the moon, it can hold an initially tenuous hydrostatic atmosphere. Its structure is well described by the classical set of  1D stellar structure equations except for the nuclear energy release term that has to be replaced by heating due to infalling planetesimals. To solve the structure equations, boundary values must be specified. In early models the outer boundary conditions, the background nebula values, were often taken to be constant, and opacities similar to the one of the interstellar medium were used.  

The calculations presented in \citet{CMpollacketal1996} treated for the first time the accretion rates of gas and solids in a self-consistent way. Their calculations show three distinct phases  (fig. \ref{fig:CMJ1}):
\begin{figure}
\begin{center}
\includegraphics[scale=0.4]{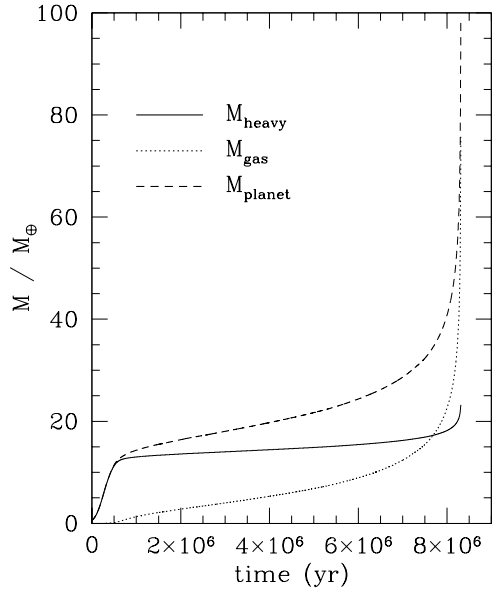}
\caption{Time evolution of the mass of accreted planetesimals (solid line), the mass of accreted gas (dotted line) and total mass of the planet (dashed line), for initial conditions equivalent to the preferred model for the formation of Jupiter in \citet{CMpollacketal1996} (from \citeauthor{CMalibertetal2005} \citeyear{CMalibertetal2005}). }\label{fig:CMJ1}
\end{center}
\end{figure}
Phase I ($t \leq $ 0.5 Myr) is characterized by a rapid build-up of  the core. It ends when all the planetesimals in the core's initial feeding zone have been accreted. During phase II ($0.5\leq t \leq 7.5 $ Myr), the core is capable of extending its feeding zone by slowly accreting some surrounding gas. An increased core mass leads to a deeper potential well which leads to a contraction of the envelope which in turns leads to additional gas accretion from the surrounding disk and so on until the systems runs away and enters phase III ($t \geq$ 7.5 Myr). This phase starts at the critical mass which is reached when the mass of the core and the envelope become roughly equal. Runaway accretion occurs because in this regime the radiative losses from the envelope can no longer be compensated for by the accretional luminosity from the planetesimals. As a result, there is no equilibrium anymore and the envelope begins to contract on much shorter timescales.  This contraction increases the gas accretion rate, which in turn increases the energy loses and the process runs away. The existence of such a critical mass is intrinsic to the core-envelope model and does not depend upon the detail of the input physics \citep{CMstevenson1982}. The critical core mass is usually of the order of 5 to 20 Earth masses \citep{CMpapaloizouterquem1999}. In the runaway phase, the gas accretion rate is limited either by the planet itself (its ability to radiate away the gravitational energy) or by how much gas the disk can supply.
 
The baseline core-accretion formation model has many appealing features, producing in a nebula with a surface density about four times the ``minimum-mass'' solar nebula a Jupiter like planet with an internal composition compatible to what is inferred from internal structure models (though the uncertainties in these models also allow a Jupiter without solid core \citep{CMsaumonguillot2004}).  However, the timescale to form the planet (8 Myrs) has always been considered as uncomfortably long compared to observationally derived lifetimes of protoplanetary disks \citep*{CMhaischetal2001}. Higher surface densities lead to significantly shorter formation timescales, which means that the baseline core accretion process is not intrinsically slow, but at the price that the resulting final content of heavy elements is very high \citep{CMpollacketal1996}. This ``timescale problem'' finally lead to the hypothesis that  another, faster formation mechanism might be needed for giant gaseous planets \citep{CMboss1997}.

\subsection{Improved and Extended Formation Models}
Since \citet{CMpollacketal1996}, the core accretion model has been significantly improved and extended, so that this ``formation timescale problem'' is no longer a problem. Improved models use more accurate equations of state \citep{CMsaumonguillot2004}, more adequate solid accretion rates \citep{CMfortieretal2007}, take into account local density patterns in the disk \citep{CMklahrbodenheimer2006} and the probable difference in opacity generated by grains in a planetary envelope as opposed to the interstellar medium \citep{CMikomaetal2000}. Indeed, as argued by \citet{CMpodolak2003}, grains entering the protoplanetary envelope may coagulate and settle out quickly into warmer regions where they are destroyed. The resulting lower opacity speeds up significantly the gas accretion timescale, while still fulfilling the observational constraint on the abundance of heavy element \citep{CMhubickyjetal2005}.
 
Extended core accretion models have included new physical mechanisms: A concurrent calculation of the evolution of the protoplanetary disk, and most importantly, planetary migration.  The discovery of numerous Hot Jupiters has forced upon us the necessity of planetary migration as in-situ formation of these objects is beyond the capability of any known formation theory.  As planet formation, disk evolution and migration occur all on similar timescales, it is necessary to treat these processes in a self consistent, coupled manner \citep*{CMalibertetal2004}.
Extending the standard core accretion model by these mechanisms not only leads to a natural explanation of the ``Hot'' planets \citep{CMalibertetal2006}, but also solves the ``timescale problem'':   For the same initial conditions that lead to runaway growth in the in situ case after 30 Myr, including concurrent disk evolution and migration leads to the formation of a Jupiter-like planet at 5.5 AU that has an internal composition compatible with internal structure models in just $\sim$1 Myr starting with an embryo of 0.6 $M_{\oplus}$ at 8 AU \citep{CMalibertetal2005}.   The reason for this speed up is that owing to migration, the planet's feeding zone is never as severely depleted as in \citet{CMpollacketal1996} and the lengthy phase II is skipped. Instead, the planet always migrates into regions of the disk where fresh planetesimals are available. 

\section{Population Synthesis}
The improvements of the core accretion model in the past few years have made possible to account for the wealth of detailed in situ and remote measurements that exist for Jupiter and Saturn  \citep{CMalibertetal2005b}, showing that the core accretion paradigm has reached a significant level of maturity. In the same time,  on the observational side immense progress has been made in both finding and characterizing new, extrasolar examples of the end products of the planetary formation process, but also the initial conditions for this process, i.e. the properties of protoplanetary disks. These two developments have opened a new, fruitful possibility of bringing together observation and theory \citep{CMidalin2004a,CMidalin2004b,CMidalin2005}: The synthesis of populations of (extrasolar) planets from a core accretion model, which are then compared to the actually observed population.   

In this section we present some of the results of population synthesis calculations we have performed with our extended core accretion model. These simulations will be presented in a series of papers (Mordasini et al. in prep., Alibert et al. in prep.).  Here, we can discuss only a very small fraction of our results, limiting ourselves to a host star mass of 1 $M_{\rm sun}$.

\subsection{Probability Distributions}
Population synthesis by means of a Monte Carlo approach rests on our ability to generate an appropriate set of initial conditions. To achieve this, the parameters of the model specifying these initial conditions have to be drawn following suitable probability distributions. In our work, these probability distributions have been derived as much a possible from observations.  

The first Monte Carlo variable is the gas to dust ratio that we relate to the observed (stellar) metallicity, using the distribution of [Fe/H] of the CORALIE planet search sample  \citep{CMsantosetal2003}. The second variable is the initial gas surface density, which is constrained by the observed distribution of disk masses in $\rho$ Ophiuchi \citep{CMbeckwithsargent1996}. The third variable is the photoevaporation rate which controls, together with the viscous evolution, the disk lifetime. We have adjusted the distribution of the photoevaporation rates so that the fraction of remaining disk as a function of time is the same as observed by \citet{CMhaischetal2001}.  The fourth Monte Carlo variable describes the initial position of the planetary seed we introduce in the protoplanetary disk. Since there are no observational constraints, we follow \citet{CMidalin2004a} and distribute our seeds with an uniform distribution in log of the semi-major axis in the regions of the disk where the local isolation mass exceeds the seeds mass (0.6 $M_{\oplus}$). Note that due to the relatively large seed mass and the other model restrictions, our synthetic planet population is incomplete in the mass domain of a few $M_{\oplus}$. In other words, our model is not a terrestrial planet formation model and should be completed appropriately before discussing this mass range.

\subsection{Detection Biases}
In order to compare to actual observations, once a synthetic population has been calculated we have to sort out the sub-population of planets which could have been detected by a given technique. For the radial velocity method, the detection probability depends to first order on the induced velocity amplitude and the instrumental accuracy. However, a large number of other quantities also affect the detection probability such as the magnitude of the star, its rotation rate, the actual measurement schedule, jitter, etc.  To compute the actual observational detection bias, we use  the method of \citet{CMnaefetal2005}  which takes all these factors into account. 

\subsection{Results}
Here we limit ourselves to the description of our nominal planet population model and restrict the host stars to  solar-type stars.  The most important parameters characterizing the nominal case are a type I migration reduction factor $f_{I}=0.01$ and a disk viscosity parameter $\alpha=0.01$  (see \citet{CMalibertetal2005} for an explanation of the quantities). To fix these parameters we compared the sub-population of detectable synthetic planets with known extrasolar planets orbiting single solar-type stars on low eccentricities orbits. We used Kolmogorov-Smirnov (KS) tests to assess the statistical significance of the synthetic mass-distance distribution (2D KS test, \citeauthor{CMpressetal1992} \citeyear{CMpressetal1992}), of the projected mass, the semimajor axis and the [Fe/H] distributions (1D KS tests), we also checked that we can reproduce the observed frequency of Hot Jupiters as well as the strong correlation of [Fe/H] and the detection probability \citep{CMfischervalenti2005}.  Fulfilling all these observational constraints \textit{simultaneously}  with a single synthetic population turned out to be a delicate task, as many combinations of parameters resulted in populations incompatible with the observational data.

\subsubsection{Mass-Distance Diagram} 
\begin{figure}
     \begin{minipage}{0.49\textwidth}
      \centering
       \includegraphics[width=\textwidth]{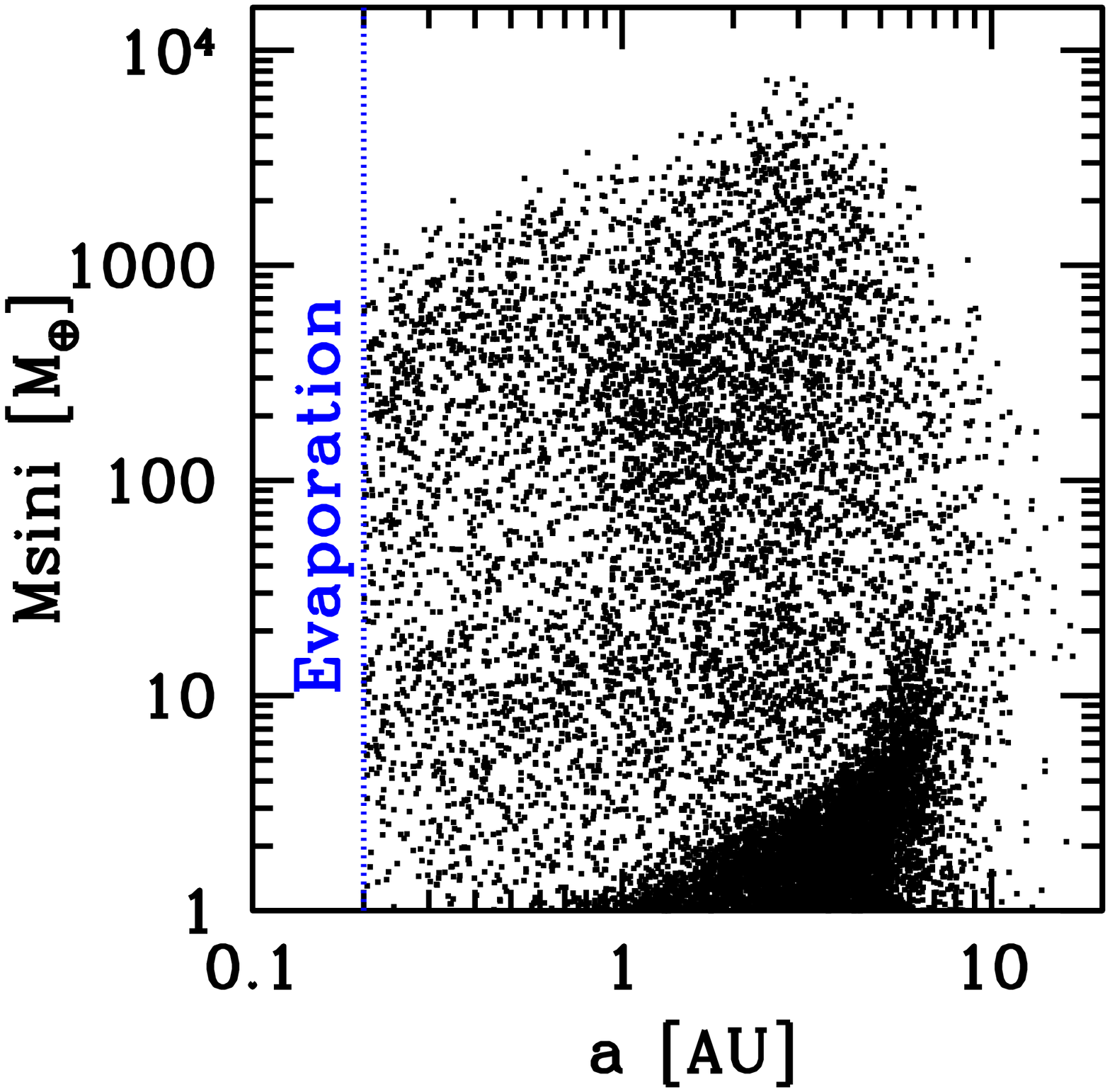}
     \end{minipage}\hfill
     \begin{minipage}{0.49\textwidth}
      \centering
       \includegraphics[width=\textwidth]{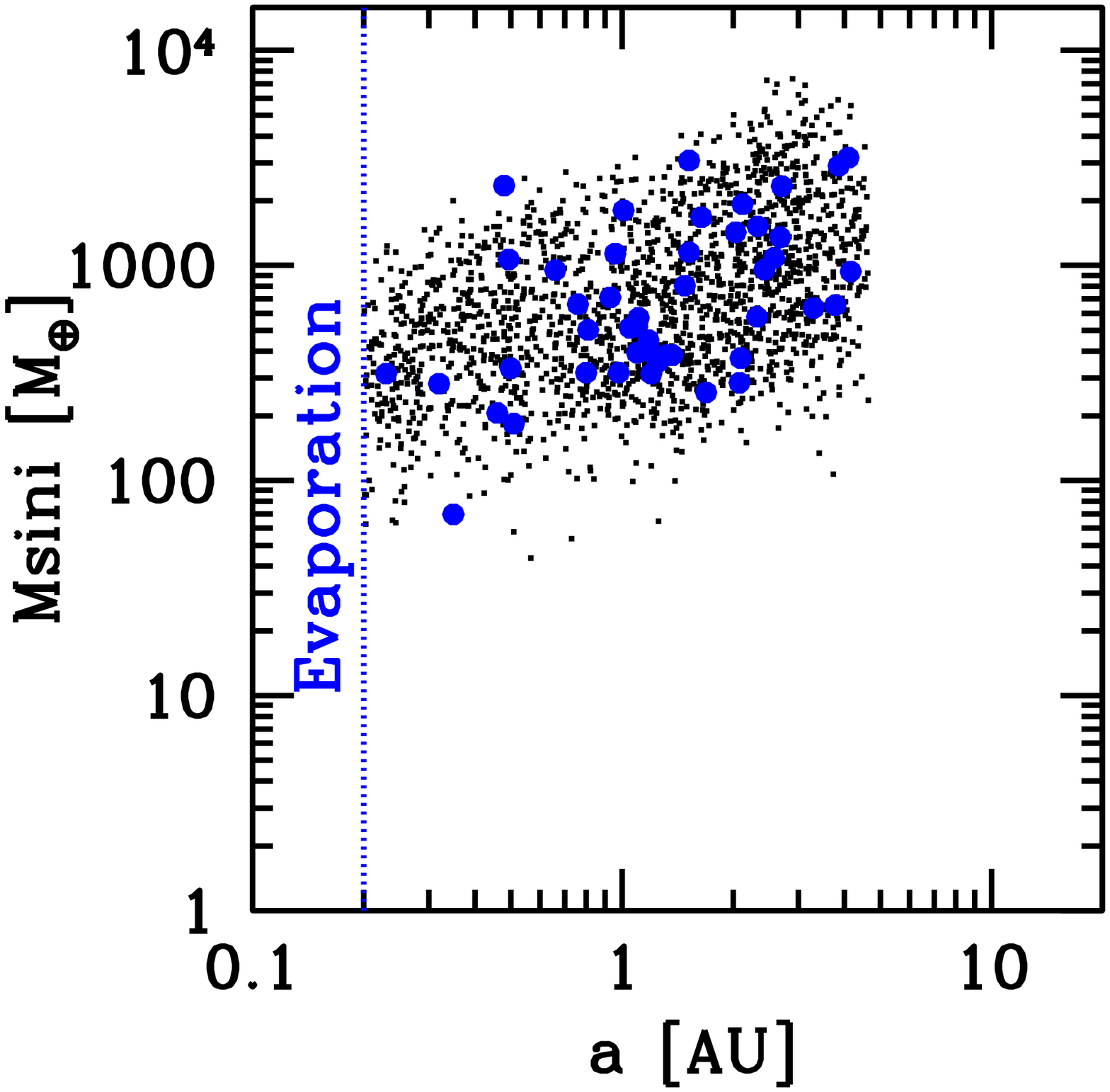}
     \end{minipage}
     \caption{Left: Minimum mass versus semimajor axis for all synthetic planets of the nominal population.  Right: The sub-population of  synthetic planets detectable by a radial velocity survey with an instrumental accuracy of 10 m/s and duration of 10 years. Real extrasolar planets  are indicated by large dots. Near the star subsequent evaporation of planets could be significant, making a direct comparison with the observations difficult.}
     \label{fig:CMaM}
\end{figure}

The KS result for the two dimensional distribution in the mass-distance plane is the observational constraint that we weighted highest, as it is of similar importance to planetary formation as the Hertzsprung-Russell diagram for stellar evolution \citep{CMidalin2004a}.  In fig. \ref{fig:CMaM}, the left panel shows the $a-M \sin i$ of all synthetic planets.
A close inspection of this figure shows that synthetic planets are not uniformly distributed but that some regions are more populated than others. The most obvious one is the vast population of \textit{failed cores}, which are planets with masses of about $1\leq M/M_{\oplus} \leq 10$ seen at the lower mass boundary in the plot. At semimajor axes of about 0.8 to 4 AU and a mass of about $100\leq M/M_{\oplus} \leq 1000$ a concentration of giant gaseous planets can be seen. This is the \textit{main clump} as it corresponds to the location of giant planets for the most common initial condition. These planets typically started at an initial semimajor axis just outside the iceline \citep{CMidalin2004a}. For $a\leq5$ AU, and $5\leq M/M_{\oplus}\leq25$, a \textit{horizontal bar} is visible which corresponds to subcritical, core dominated, planets. These planets migrated inwards, mainly by type II migration, while only slowly accreting mass. Some of them grew massive enough to become super-critical thus leaving the horizontal bar towards the giant planet region. Others, never reached the critical mass and kept migrating inward building-up the population of Hot Neptunes.  Above the horizontal bar, and to the left of the main clump, at masses between 25 to 200 $M_{\oplus}$, there is a region with somewhat less planets. This is the equivalent to the \textit{planetary desert} described by \citet{CMidalin2004a}. Compared to their calculations, the planetary desert in our calculations is certainly less pronounced. We note that this shallower planetary desert is not simply due to a $\sin i$ effect.

The right panel of fig. \ref{fig:CMaM} shows the sub-population of  synthetic planets detectable by a radial velocity survey with an instrumental accuracy of 10 m/s and a duration of 10 years. It is obvious that at that precision, only a small fraction (6.1\%) of all synthetic planets can be detected. In the figure, the comparison sample of actual extrasolar planets is also plotted. Comparing these two populations in a 2D KS test leads to a significance of 53\% that both come from the same parent distribution. Comparing the $M \sin i$ and the semimajor axis distribution separately in 1D KS tests leads to a significance of 90\% and 33\%, respectively.  The better KS result for $M \sin i$ than for $a$ is attributed to the fact that in our models, migration is computed in a cruder approximation than mass accretion.  

\subsubsection{Planetary IMF}
\begin{figure}
\begin{center}
\includegraphics[scale=0.65]{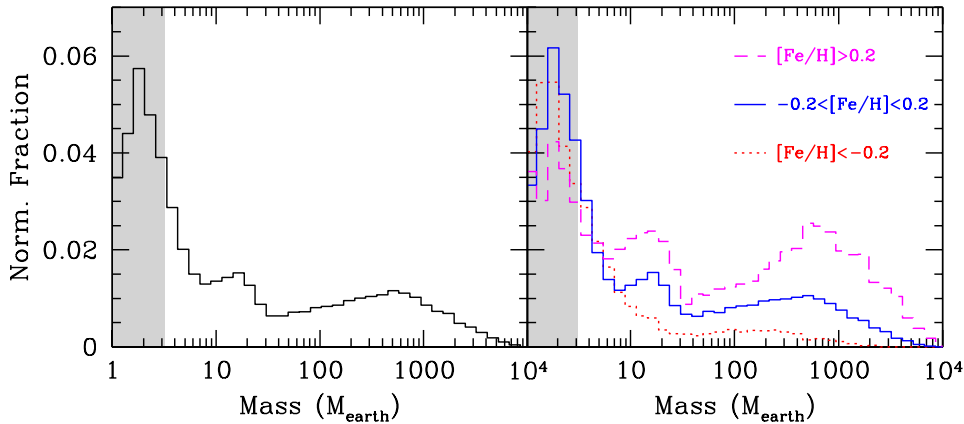}
\caption{Initial mass function of all planets of the synthetic population around G type stars. In the right panel, the population has been split in a low, medium and high metallicity bin. The region of a few $M_{\oplus}$ has been shaded as the model is incomplete there.}\label{fig:CMmasshisto} 
\end{center}
\end{figure}

At a RV precision of 10 m/s, our knowledge of the planetary mass mass function was limited to the giant planets. At 1 m/s, the precision reached by HARPS, Neptune mass planets became detectable. In future, at a precision of  0.1 m/s, we will descend into the terrestrial mass regime.  It is therefore of interest to study the underlying, unbiased IMF (fig. \ref{fig:CMmasshisto}).
From the left panel, it is seen that the IMF has a quite complex structure. Starting at the large mass end, we note that  core accretion is able to form planets that can, at least if the presence of the core is not important, ignite deuterium burning. However, such planets are rare. At about 500 $M_{\oplus}$, the IMF has a local maximum, followed by a local minimum at $\sim40$ $M_{\oplus}$.   A small bump occurs in the Neptune mass domain due to the horizontal bar. At around 6 $M_{\oplus}$, the IMF finally starts to raise rapidly. As our model is incomplete for these very low masses, quantitative predictions should be regarded with caution here. Qualitatively, a strong raise of the IMF at such masses is however nevertheless expected, as  it is simply a consequence of the fact that very often the conditions in the protoplanetary nebula are such that they don't allow the formation of a giant planet.

In the right panel, the population was split in a low, medium and high metallicity bin. One can see that the IMF is clearly metallicity dependent, with metal rich systems producing more massive objects, and metal poor more small bodies.  The distributions cross at around 6 $M_{\oplus}$. This metallicity dependent IMF explains why radial velocity technique based planet searches, which are biased towards large masses, have found planets preferentially orbiting metal rich stars.  

\section{Conclusion}
The core accretion paradigm explains in an unified way the formation of giant and terrestrial planets, so that there is no need for a special mechanism for giant planets.  Since the first core accretion models, significant improvement and extensions were made. Such improved and extended core accretion models can form giant planets well within observed disk lifetimes, so that there is no need for a faster formation mechanism.  They have also reached a degree of maturity that allows quantitative tests with observations, of both the giant planets of our own Solar System, and, by means of population synthesis, of the extrasolar planet population. In the latter case, the whole population of detected planets can be used to constrain the models, which excludes model fine tuning for a specific case, and fully exploits the observational investment. 
As shown by statistical tests,  extended core accretion models can reproduce many observed properties and correlations in the extrasolar planet population in a quantitative significant way with one synthetic population at one time. This means that accretion models can now be used to predict future observations, so that theory can feed back on the design of future instruments

\acknowledgements We thank Isabelle Baraffe, Gilles Chabrier and Stephane Udry for useful discussions.  This work was supported in part by the Swiss National Science Foundation. Computations were made on the ISIS and ISISII clusters at the University of Bern.

\end{document}